\documentclass{camera}
\usepackage{graphicx,bibspacing}  

\begin{document}

%
\title{Finite lifetime effects on the photon production from a quark-gluon plasma}

%
\author{F. Michler$^{1}$, B. Schenke$^{2}$ \and C. Greiner$^{1}$}

%
\organization{$^{1}$Institut f\"ur Theoretische Physik \\
              Johann Wolfgang Goethe - Universit\"at Frankfurt \\
              Max-von-Laue-Stra\ss{}e~1,
              D-60438 Frankfurt am Main, Germany, \\
              \mbox{ } \\
              $^{2}$Department of Physics, McGill University, \\
              Montreal, Quebec, H3A\,2T8, Canada}

\maketitle

\begin{abstract}
We use the real-time Keldysh formalism to investigate finite lifetime effects on the photon emission from a quark-gluon plasma (QGP). We provide an ansatz which eliminates the divergent contribution from the vacuum polarization and renders the photon spectrum UV-finite if the time evolution of the QGP is described in a suitable manner.
\end{abstract}

%
The quark-gluon plasma (QGP) created during heavy-ion collisions can only be accessed indirectly via experimental signatures such as hard and electromagnetic probes. Besides the role of memory effects in time evolution (see \textit{e.g.} \cite{SG:2005,Michler:2009dy}) it is also of particular interest within the framework of non-equilibrium quantum field theory how these probes are affected by the finite lifetime of the QGP itself.

For photons, this question has first been addressed by Boyanovsky \textit{et al.} \cite{Wang:2000pv,Wang:2001xh}. The main result has been the contribution of first order processes which are kinematically forbidden in equilibrium. The spectra from these processes were found to decay algebraically for $k>1.5$ GeV and thus dominate over higher order equilibrium contributions in that range. Two major problems arose in the above investigations, namely the divergent contribution from the vacuum polarization and the non-integrability of the remaining contributions in the ultraviolet domain \cite{Boyanovsky:2003qm}. The topic was then touched on by Fraga \textit{et al.} \cite{Fraga:2003sn} where it was claimed that the divergent vacuum contribution is unphysical and thus requires an appropriate renormalization technique. It was concluded that the ansatz used in \cite{Boyanovsky:2003qm} is inadequate as it produces the mentioned problems.

\begin{figure}[htb]
 \center
 \includegraphics[height=4.5cm]{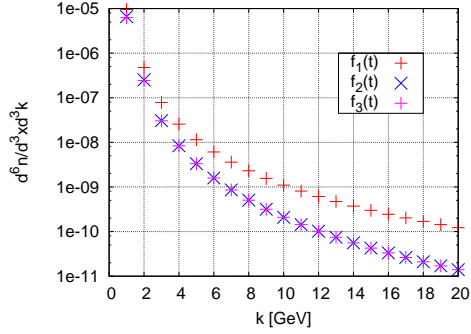}
 \caption{The photon spectrum decays as $\sim 1/k^{3}$ independently from $f(t)$.}
 \label{fig:scenario_1}
\end{figure}

We provide an ansatz that eliminates the divergent contribution from the vacuum polarization. For the scenario of a heavy-ion collision, it also renders the resulting photon spectrum UV-finite if the time evolution is described in a suitable manner.\\
\\
The photon production rate from a homogeneous but non-stationary emitting system reads \cite{Michler:2009dy,Serreau:2003wr}:
\begin{equation}
 \label{eq:photon_rate}
 k\frac{d^{7}n(t)}{d^{4}xd^{3}k} = \frac{1}{(2\pi)^{3}}\mbox{Re}\left\lbrace\int_{-\infty}^{t}du\mbox{ }
                                   i\Pi^{<}_{T}(\vec{k},t,u)e^{ik(t-u)}\right\rbrace
\end{equation}
As in \cite{Boyanovsky:2003qm}, we use the one-loop approximation for the photon self energy $i\Pi^{<}_{T}(\vec{k},t,u)$ including the processes of first order in $\alpha_{e}$. We model the finite lifetime of the QGP via time dependent occupation numbers of the quarks
\begin{equation}
 n_{F}(E) \rightarrow n_{F}(E,t) = f(t)n_{F}(E)\mbox{ ,}
\end{equation}
and couple the time evolution to the interaction vertices by replacing the occupation numbers and the number of holes in $i\Pi^{<}_{T}(\vec{k},t,u)$ by their geometric mean from the two times $t$ and $u$. If we now decompose $i\Pi^{<}_{T}(\vec{k},t,u)$ into the vacuum polarization and the medium contribution
\begin{equation}
 \label{eq:polmomentum}
 i\Pi^{<}_{T}(\vec{k},t,u) = i\Pi^{<}_{T,0}(\vec{k},t-u)+i\Pi^{<}_{T,M}(\vec{k},t,u) \, , \
\end{equation}
and insert (\ref{eq:polmomentum}) into (\ref{eq:photon_rate}), we see that the vacuum polarization is evaluated on-shell and thus does not contribute. Therefore, the divergence associated with it does not show up in the photon yield. As the occupation numbers are time dependent, the medium part of the photon self energy is evaluated off-shell which makes the contribution of first order processes possible.\\
\\
For our numerical investigations, we consider a QGP with a temperature of $T=0.3$ GeV. It is 'switched on' over a timescale of $\tau_{F}=1$ fm/c which is modeled by three different switching functions $f(t)$.
\begin{eqnarray}
 f_{1}(t) & = & \Theta(t) \nonumber \\
 f_{2}(t) & = & \Theta(-t)e^{t/t_{F}}+\Theta(t) \nonumber \\
 f_{3}(t) & = & \frac{1}{2}\Theta(-t)e^{t/t_{F}}+\Theta(t)\left[1-\frac{1}{2}e^{-t/t_{F}}\right] \nonumber
\end{eqnarray}
Figure \ref{fig:scenario_1} shows that the resulting photon spectrum decays as $\sim 1/k^{3}$ for all switching functions $f(t)$ and is thus not integrable.

This behaviour can be understood as follows. The dominant contributions to the photon yield are Bremsstrahlung and a negative contribution from Pauli Blocking of the pair creation process. Both of them behave as $\sim 1/k^3$ for large $k$ independently of $f(t)$ \cite{Michler:2009}. If we turn from an instantaneous to a smoother switching, the Bremsstrahlung contribution essentially halves in value whereas the Pauli Blocking contribution is left unchanged \cite{Michler:2009}. This accounts for the slightly steeper decay in this case.\\
\\
The problem with the UV-integrability can be circumvented if the QGP is also switched off again as to mimique a heavy-ion collision. The photon yield is assumed to be observed at $t\rightarrow\infty$. We adopt a creation and hadronization time of $\tau_{F}=\tau_{H}=1$ fm/c and a lifetime of $\tau_{L}=4$ fm/c. Again different switching functions $g(t)$ are compared.
\begin{eqnarray}
 g_{1}(t) & = & f_{1}(t)\cdot\Theta(\tau_{L}-t) \nonumber \\
 g_{2}(t) & = & f_{2}(t)\cdot\left[\Theta(\tau_{L}-t)+\Theta(t-\tau_{L})e^{-(t-\tau_{L})/t_{H}}\right] \nonumber \\
 g_{3}(t) & = & f_{2}(t)\cdot\left[\Theta(\tau_{L}-t)\left(1-\frac{1}{2}e^{(t-\tau_{L})/t_{H}}\right)
                         +\frac{1}{2}\Theta(t-\tau_{L})e^{-(t-\tau_{L})/t_{H}}\right] \nonumber
\end{eqnarray}

\begin{figure}[htb]
 \center
 \includegraphics[height=4.5cm]{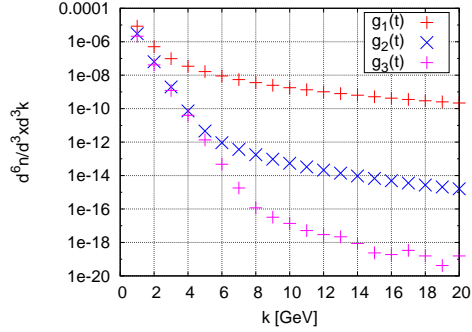}
 \caption{The decay of the photon spectrum is highly sensitive to $g(t)$.}
 \label{fig:scenario_2}
\end{figure}

In this case, the negative Pauli Blocking contribution vanishes \cite{Michler:2009}. One can furthermore infer from fig. (\ref{fig:scenario_2}) that the resulting photon spectrum strongly depends on the considered switching function $g(t)$. For an instantaneous switching, it still decays as $\sim 1/k^{3}$ for large $k$. But if we consider an exponential ($g_{2}(t)$) or an even smoother switching ($g_{3}(t)$) which represents a more realistic scenario, the spectrum is suppressed to $\sim 1/k^{5}$ and $\sim 1/k^{7}$, respectively, and thus rendered UV-finite.

In summary, we have presented an ansatz which eliminates the divergent photon yield from the vacuum polarization and partially solves the problem with the UV-integrability. As the next step, we have to revisit it in a way that it also leads to UV-finite photon spectra for the general case. We also will consider the treatment of possible infrared singularities.\\
\\
F. M. gratefully acknowledges financial support by the Helmholtz Research School for Quark Matter Studies (H-QM) and by the Helmholtz Graduate School for Hadron and Ion Research (HGS-HIRe for FAIR). B. S. gratefully acknowledges a Richard H. Tomlinson grant by McGill University and support by the Natural Sciences and Engineering Research Council of Canada. This work was (financially) supported by the Helmholtz International Center for FAIR within the framework of the LOEWE program (Landesoffensive zur Entwicklung Wissenschaftlich-\"Okonomischer Exzellenz) launched by the State of Hesse.



%
\bibliography{contribution_michler}
\setlength{\bibspacing}{\baselineskip}
\bibliographystyle{phjcp}

\end{document}